\documentclass[12pt]{article}
\usepackage[dvips]{graphicx}
\usepackage{pdproc}

  \makeatletter 
  \def\@cite#1{[#1]} 
  \makeatother    
\begin{document}

\renewcommand{\thefootnote}{\alph{footnote}}

\title{
 Non-Abelian Walls and Vortices in Supersymmetric Theories 
}

\author{ 
Youichi~Isozumi, 
Muneto~Nitta, 
 Keisuke~Ohashi, 
 and 
Norisuke~Sakai\footnote
{Speaker at the conference}
}

\address{ 
Department of Physics, Tokyo Institute of Technology \\
Tokyo 152-8551, JAPAN 
}

\abstract{
We review recent results on solitons in supersymmetric 
(SUSY) non-Abelian gauge theories, focusing on our papers: 
hep-th/0405194, hep-th/0405129, and hep-th/0404198. 
We construct the BPS 
multi-wall solutions in 
supersymmetric 
$U(N_{\rm C})$ gauge theories in five dimensions with 
$N_{\rm F}(>N_{\rm C})$ hypermultiplets in the fundamental representation. 
Exact solutions are obtained with full generic moduli 
for infinite gauge coupling. 
Total moduli space of the BPS non-Abelian walls 
is found to be the complex Grassmann manifold 
$SU(N_{\rm F}) / 
[SU(N_{\rm C})\times SU(N_{\rm F}-{N}_{\rm C}) \times U(1)]$. 
A 1/4 BPS equation is also studied which gives 
combinations of vortices, walls and monopoles. 
The full moduli space of the 1/4 BPS equation is found 
to be the space of all holomorphic maps from a complex 
plane to the wall moduli space. 
Exact solutions of the 1/4 BPS equation are also obtained 
for infinite gauge coupling. 
}

\normalsize\baselineskip=15pt

\section{Introduction}

Supersymmetric (SUSY) theories have been providing the most 
realistic models beyond the standard model~\cite{DGSW}. 
In recent years, 
much attention has been devoted to 
models with extra 
dimension~\cite{HoravaWitten}--
\cite{RandallSundrum}, where our 
four-dimensional world is realized on a topological 
defect in higher dimensional spacetime. 
It is desirable to obtain the topological defects 
as solitons in field theories. 
In constructing topological defects, supersymmetric 
theories are helpful, since partial preservation of 
SUSY automatically gives a solution of equations of 
motion~\cite{WittenOlive}. These states are called 
BPS states. 
Moreover, the resulting theory 
tend to produce an ${\cal N}=1$ 
SUSY theory on the world volume, 
which can provide realistic unified models with the 
desirable properties. 
Supersymmetry also helps to obtain stability of the 
soliton. 
The simplest of such solitons is the domain 
wall~\cite{CQR}. 
It has been a difficult problem to localize massless gauge 
bosons on the wall in five dimensions, in spite of many 
interesting proposals, especially in lower 
dimensions~\cite{DvaliShifman}--\cite{SY2}. 
Recently a model of the localized massless gauge 
bosons on the wall has 
been obtained for Abelian gauge theories 
using SUSY QED interacting with 
tensor multiplets~\cite{IOS1,IOS2}. 
Walls in non-Abelian gauge theories 
are likely to help 
obtaining non-Abelian gauge bosons localized 
on the world volume. 
We call these walls as 
non-Abelian walls. 
Non-Abelian wall solutions 
have rich structures and are also interesting 
in its own right. 

Recently we have conducted a series of study 
to obtain various BPS solutions of 
a non-Abelian gauge theory with 
eight supercharges in spacetime dimensions of five or 
less~\cite{INOS1}--\cite{INOS3}. 
More specifically, we have studied the $U(N_{\rm C})$ 
gauge theory with $N_{\rm F}(>N_{\rm C})$ flavors of 
hypermultiplets in the fundamental representation. 
To obtain discrete vacua, we consider non-degenerate 
masses for hypermultiplets, 
and the Fayet-Iliopoulos (FI) parameter is introduced~\cite{ANS}. 
We have obtained BPS multi-wall solutions as $1/2$ BPS 
solutions, and various combinations of walls, vortices 
and monopoles in the Higgs phase as $1/4$ BPS solutions. 
By taking the limit of infinite gauge coupling, 
we have obtained exact BPS multi-wall solutions with generic 
moduli parameters covering the complete moduli space 
of walls~\cite{INOS1}. 
For a restricted class of moduli parameters called 
$U(1)$-factorizable moduli, 
we have obtained exact BPS multi-wall solutions for 
certain values of finite gauge coupling~\cite{INOS3}. 

Since the BPS equations for the hypermultiplets $H^i$ 
involve the extra-dimensional component 
gauge field $W_y$ and the vector multiplet scalar 
as a function of the extra-dimensional coordinate $y$, 
it can be solved in terms of a $GL(N_{\rm C}, {\bf C})$ 
gauge transformation matrix $S(y)$ as a function of 
the extra-dimensional coordinate $y$. 
Apart from the additional $y$ dependence due to the 
mass of each flavor, the hypermultiplet is 
thus given by a product~\cite{INOS1} of $S(y)$ and 
$N_{\rm C}\times N_{\rm F}$ 
constant moduli matrices $H_0^i$. 
The BPS equation for the vector multiplet determines 
$S(y)$ once $H_0$ is given. 
Since the hypermultiplet depends only on the product 
$S^{-1}H_0$, this description has a redundancy of 
a global symmetry $GL(N_{\rm C}, {\bf C})$ 
acting on $(S, H_0^i)$, which we call the 
world-volume symmetry. 
Therefore the moduli space of $1/2$ BPS domain walls 
is given by 
a compact complex manifold, the Grassmann manifold~\cite{INOS1} 
$G_{N_{\rm F},N_{\rm C}}\equiv {SU(N_{\rm F}) \over 
SU(N_{\rm C})\times SU(N_{\rm F}-N_{\rm C}) 
\times U(1)}$. 
One should note that this is 
the total moduli space of the multi-wall solutions 
including all the topological sectors. 
If we take  a multi-wall configuration and 
let one of the wall going to infinity, we obtain a configuration 
with one less wall. 
Therefore configurations with smaller number of domain 
walls appear as boundaries of the moduli space. 
If we add all these different topological sectors, the total 
moduli space becomes the compact complex manifold 
$G_{N_{\rm F},N_{\rm C}}$. 

The world-volume symmetry turned out to be very 
useful and will eventually be promoted to a local 
gauge symmetry when we consider effective theories 
on the world volume of walls.

If we consider a vortex perpendicular to a wall, 
it preserves a different combination of four 
supercharges compared to 
the wall~\cite{vortex}. 
We have found that the coexistence of the vortex 
and the wall can be realized as a $1/4$ 
BPS configuration~\cite{INOS2}. 
We also found that this $1/4$ BPS equation admits also 
monopoles in the Higgs phase which were found 
recently~\cite{mono-Higgs}. 
We have obtained the exact 
solutions in the limit of infinite gauge coupling. 
We also identified  the moduli space of the $1/4$ BPS 
solutions to be the space of 
all the holomorphic maps from the complex plane to 
the wall moduli space~\cite{INOS2}, 
the Grassmann manifold 
$G_{N_{\rm F},N_{\rm C}}$.

\section{SUSY Vacua of $U(N_{\rm C})$ Gauge Theories }
\label{sc:model-vacua-BPSeq}

The minimum number of supercharges in five dimensions 
is eight. 
Building blocks of field theories with eight 
supercharges are 
vector multiplets and hypermultiplets. 
Since wall solutions require discrete vacua, 
we consider a gauge group with 
$U(1)$ factors besides semi-simple gauge group~\cite{ANS}. 
As the most natural non-Abelian gauge group, 
we choose $U_{\rm G}(N_{\rm C})$ gauge group. 
We denote the gauge group suffix 
and flavor group suffix in our fundamental theory 
by the uppercase letter G and F, respectively. 
Let us consider a five-dimensional SUSY model with minimal 
kinetic terms for vector and hypermultiplets. 
Physical bosonic fields of the vector multiplet are 
gauge field $W_M$, and a scalar $\Sigma$ in the adjoint 
representation of the $U_{\rm G}(N_{\rm C})$ gauge group. 
The fields in the vector multiplet such as $\Sigma$ 
are expressed by an $N_{\rm C}\times N_{\rm C}$ matrix. 
Physical bosonic field of the hypermultiplet is 
an $SU(2)_R$ doublet of complex scalars $H^{i}, i=1,2$ 
which are assigned to the fundamental representation 
of $U_{\rm G}(N_{\rm C})$. 
We consider $N_{\rm F}$ flavors of hypermultiplets and 
combine them into an $N_{\rm C}\times N_{\rm F}$ 
matrix denoted as $H^i\equiv(H^{irA}) $ with 
color $r=1, \cdots, N_{\rm C}$ and flavor 
$A=1, \cdots, N_{\rm F}$ indices. 
We denote space-time indices by 
$M,N, \cdots=0,1,2,3,4$ with the metric 
$\eta_{MN}={\rm diag}(+1,-1,-1,-1,-1)$. 
For simplicity, we take the same gauge coupling $g$ for 
$U(1)$ and $SU(N_{\rm C})$ factors. 
The eight supercharges 
allow only a few parameters in our model: 
gauge coupling constant $g$, 
the masses of $A$-th hypermultiplet $m_A$, and 
the FI parameter $c$ for the $U(1)_{\rm G}$ gauge group. 
Then the bosonic part of our Lagrangian reads 
(after eliminating the auxiliary fields) 
\begin{equation}
{\cal L} 
=
-\frac{1}{2g^2}{\rm Tr}\left[ F_{MN}F^{MN}\right]
+\frac{1}{g^2}{\rm Tr}
\left[{\cal D}_M \Sigma {\cal D}^M \Sigma \right]
+{\rm Tr}
\left[ {\cal D}^M H^i ({\cal D}_M H^i)^\dagger \right] -V,
\end{equation}
\begin{eqnarray}
V
&\!\!\!=&\!\!\! 
\frac{g^2}{4}
{\rm Tr}
\left[
\left(
H^{1}  H^{1\dagger}  - H^{2} H^{2\dagger} 
- c\mathbf{1}_{N_{\rm C}}
\right)^2 +
4H^2H^{1\dagger} H^1H^{2\dagger} 
\right] 
\nonumber\\
&
&
+{\rm Tr}\left[
 (\Sigma H^i - H^i M) 
 (\Sigma H^i - H^i M)^\dagger 
 \right],
\end{eqnarray}
where the mass matrix is defined as 
$(M)^A{}_B\equiv m_A\delta ^A{}_B$, 
and the covariant derivatives as 
${\cal D}_M \Sigma = \partial_M \Sigma + i[ W_M , \Sigma ]$, 
${\cal D}_M H^{i}=(\partial_M + iW_M)H^{i} $, 
and field strength as 
$F_{MN}=\frac{1}{i}[{\cal D}_M , {\cal D}_N]
=\partial_M W_N -\partial_N W_M + i[W_M, W_N]$. 
We assume non-degenerate 
mass parameters $m_A$ and choose the order of the 
mass parameters as $m_A > m_{A+1}$ for all $A$. 
Then the flavor symmetry reduces to 
$ G_{\rm F} = U(1)_{\rm F}^{N_{\rm F}-1}$, 
where 
$U(1)_{\rm F}$ corresponding to the common phase is 
gauged by $U(1)_{\rm G}$ local gauge symmetry. 

\begin{figure}[thb]
\begin{center}
\includegraphics[width=7cm,clip]{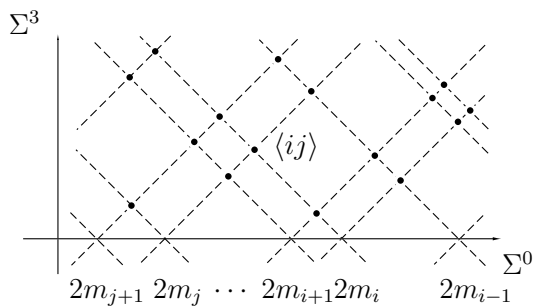}
\end{center}
\caption{
Diagrammatic representation of vacua for $N_{\rm C}=2$ case. 
Dashed lines are defined by $\Sigma^0+\Sigma^3 =2m_A$
and $\Sigma^0 - \Sigma^3 =2m_A$. 
Vacua are given as intersection 
points of these lines except for $\Sigma ^3=0$, 
because of $H^2 H^{1\dagger}=0$ in Eq.~(\ref{D-term-cond}). 
}
\label{su2nf}
\end{figure}

SUSY vacua are realized at vanishing vacuum 
energy, which requires 
both contributions from vector and hypermultiplets 
to vanish. 
Conditions of vanishing contribution from vector 
multiplet read 
\begin{eqnarray}
H^{1}  H^{1\dagger}  - H^{2} H^{2\dagger} 
=c\mathbf{1}_{N_{\rm C}},
\quad 
 H^2 H^{1\dagger}= 0.
\label{D-term-cond}
\end{eqnarray}
Requiring the vanishing contribution to vacuum energy 
from hypermultiplets 
gives the SUSY condition for hypermultiplets as 
\begin{eqnarray}
\Sigma H^i - H^i M
=0. 
\label{F-term-cond++}
\end{eqnarray}
By local gauge transformations of $G$, we can always 
choose the vector multiplet scalar to be diagonal 
$\Sigma ={\rm diag}(\Sigma_1, \Sigma_2, \cdots, \Sigma_{N_C})$. 
Then Eq.~(\ref{F-term-cond++}) becomes diagonal 
: $(\Sigma_s -m_A)  H^{isA}=0$, where $r$ and $A$ 
denote color and flavor indices. 
The condition $ H^2 H^{1\dagger}= {0}$ in 
Eq.(\ref{D-term-cond}) 
is best satisfied by $H^2 = 0$, because of $c >0$. 
Eq.(\ref{D-term-cond}) demands 
that at least some of $ H^{isA}$ have to be nonvanishing. 
Then Eq.(\ref{F-term-cond++}) requires the corresponding 
vector multiplet scalar to be nonvanishing $\Sigma_s=m_A$. 
Since we assume non-degenerate masses for hypermultiplets, 
we find that only one flavor $A=A_r$ 
can be non-vanishing 
for each color component $r$ of hypermultiplet scalars 
$H^{irA}$ 
with 
\begin{eqnarray}
 H^{1rA}=\sqrt{c}\,\delta ^{A_r}{}_A,\quad H^{2rA}=0.
 \label{eq:hyper-vacuum}
\end{eqnarray}
Here we used global gauge transformations 
to eliminate possible phase factors. 
This is called the color-flavor locked vacuum. 
The vector multiplet scalar $\Sigma$ is 
determined as 
\begin{eqnarray}
\Sigma ={\rm diag.}(m_{A_1},\,m_{A_2},\,\cdots,\,
m_{A_{N_{\rm C}}}).
\end{eqnarray}
Now we see that $N_{\rm F}>N_{\rm C}$ is needed in order to obtain 
disconnected SUSY vacua appropriate for constructing walls. 
The vacua for $U(2)_{\rm G}$ 
gauge group are illustrated in Fig.\ref{su2nf}, 
using $\Sigma\equiv {\rm diag}.((\Sigma^0+\Sigma^3)/2, 
(\Sigma^0 - \Sigma^3)/2)$. 

We denote a SUSY vacuum specified by a set of 
non-vanishing hypermultiplet 
scalars with the flavor $\{A_r\}$ for each color 
component $r$ as 
\begin{eqnarray}
 \langle A_1\,A_2\,\cdots\,A_{N_{\rm C}}\rangle .
\end{eqnarray}    
Since global gauge transformations can exchange flavors 
$A_r$ and $A_s$ for the color component $r$ and $s$, 
respectively, 
the ordering of the flavors $A_1, \cdots, A_{N_{\rm C}}$ 
does not matter in considering only vacua: 
$\langle 123 \rangle
=\langle 213 \rangle
$. 
Thus a number of SUSY 
vacua is given by~\cite{ANS} 
\begin{eqnarray}
 {}_{N_{\rm F}}C_{N_{\rm C}} 
 = {N_{\rm F}! \over N_{\rm C}!  (N_{\rm F} - N_{\rm C})!} 
   \label{vacnumber}
\end{eqnarray} 
and 
we usually take $A_1<A_2<\cdots<A_{N_{\rm C}}$.

\section{BPS Equations for Walls }

Walls interpolate between two vacua at $y=\infty$ 
and $y=-\infty$. 
These boundary conditions at $y=\pm \infty$ 
define topological sectors. 
When we consider walls, it is often 
convenient to fix a gauge in presenting solutions. 
The gauge transformations allow us to eliminate 
all the vector multiplet scalars 
$\Sigma^{I\not\in{\cal H}}$ 
for generators outside of the Cartan 
subalgebra ${\cal H}$, and all the gauge fields 
$W_y^{I\in {\cal H}}$ in  the Cartan 
subalgebra ${\cal H}$. 
In this gauge, gauge fields $W_y^{I\not\in{\cal H}}$ 
can no longer be 
eliminated, since gauge is completely fixed. 
We usually use this gauge 
\begin{equation}
\Sigma^{I\not\in{\cal H}}=0, \qquad 
W_y^{I\in {\cal H}}=0. 
\end{equation}
If we wish, we can choose another gauge where 
the extra dimension component $W_y^I$ 
of the gauge field vanishes for all the generators. 
Then all components of 
vector multiplet scalars $\Sigma^I$ 
including those out of Cartan subalgebra become 
nontrivial. 
In that gauge, our BPS multi-wall solutions are 
expressed by nontrivial vector multiplet scalars 
$\Sigma^I$ for all the generators, instead of gauge 
fields $W_y^I$. 
Our explicit solutions show that the gauge field 
content is generically nontrivial, since we cannot 
eliminate all of  $W_y^I$ and  $\Sigma^I$ : 
either $W_y^I\not=0$ or $\Sigma^I\not=0$ in any gauge.

To obtain domain walls, we assume that 
all fields depend only on the coordinate of one extra 
dimension $x^4$ which we denote as $y$. 
We also assume the Poincar\'e invariance 
on the four-dimensional world volume of the wall, 
implying 
\begin{eqnarray}
F_{MN}(W) =0, \ \ W_\mu =0,
\label{eq:gauge-field-wall}
\end{eqnarray}
where we take $x^\mu=(x^0, x^1,x^2,x^3)$ as 
four-dimensional world-volume coordinates. 
Note that $W_y$ need not vanish. 
The Bogomol'nyi completion of the energy density 
of our system can be performed as 
\begin{eqnarray}
{\cal E}
&=&
\frac{1}{g^2}{\rm Tr}\left[( {\cal D}_y \Sigma)^2\right] 
+{\rm Tr}\left[ {\cal D}^M H^i ({\cal D}_M H^i)^\dagger \right]
+{\rm Tr}\left[
 (\Sigma H^i - H^i M) 
 (\Sigma H^i - H^i M)^\dagger 
 \right]
\nonumber\\
&
&
+
\frac{g^2}{4}
{\rm Tr}
\Big[
\left(
H^{1}  H^{1\dagger}  - H^{2} H^{2\dagger} 
- c\mathbf{1}_{N_{\rm C}}
\right)^2 +
4H^2H^{1\dagger} H^1H^{2\dagger} 
\Big] 
\nonumber\\
&=&\frac{1}{g^2}{\rm Tr}\left({\cal D}_y \Sigma -
{g^2\over 2}\left(c{\bf 1}_{N_{\rm C}}-H^1H^1{}^\dagger 
+H^2H^2{}^\dagger \right)\right)^2
+
{g^2}{\rm Tr}
\Big[
H^2H^{1\dagger} H^1H^{2\dagger} 
\Big] 
\nonumber \\
&&{}
+ {\rm Tr}\left[
({\cal D}_y H^1 + \Sigma H^1 -H^1M) 
({\cal D}_y H^1 + \Sigma H^1 -H^1M)^\dagger\right] 
\nonumber \\ && {}
+ {\rm Tr}\left[
({\cal D}_y H^2 - \Sigma H^2 +H^2M) 
({\cal D}_y H^2 - \Sigma H^2 +H^2M)^\dagger\right]  
\nonumber \\ && {}
+ c \partial_y{\rm Tr}\Sigma 
- \partial_y \left\{{\rm Tr}
\left[ 
\left(\Sigma H^1 - H^1 M\right)H^1{}^\dagger
+ \left(-\Sigma H^2 
+H^2 M\right)H^2{}^\dagger\right]\right\}. 
\label{eq:bogomolnyi}
\end{eqnarray}
Therefore we obtain a lower bound for the energy of the 
configuration. 
By saturating the complete squares, 
we obtain the BPS equation 
\begin{eqnarray}
{\cal D}_y \Sigma = 
{g^2\over 2}\left(c{\bf 1}_{N_{\rm C}}-H^1H^1{}^\dagger 
+H^2H^2{}^\dagger \right), 
\quad 
0
=
g^2 H^1H^2{}^\dagger, 
\label{BPSeq-Sigma}
\end{eqnarray}
\begin{eqnarray}
{\cal D}_y H^1 
=
-\Sigma H^1 + H^1 M,\qquad 
{\cal D}_y H^2 = \Sigma H^2 -H^2 M. 
\label{BPSeq-H}
\end{eqnarray}
The BPS saturated configuration satisfying the BPS 
Eqs.~(\ref{BPSeq-Sigma})-(\ref{BPSeq-H}) gives the 
minimum energy (per unit volume), namely the tension 
of the BPS wall as 
\begin{eqnarray}
T_{\rm w}&\!\!=\!\!&\!\!\int^{+\infty}_{-\infty}
\hspace{-1.5em}dy{\cal E}
=c 
\left[{\rm Tr}\Sigma \right]^{+\infty}_{-\infty}
\!\!=\!c \left(\sum_{k=1}^{N_{\rm C}}m_{A_k}
-\sum_{k=1}^{N_{\rm C}}m_{B_k}\right), 
\label{eq:tension}
\end{eqnarray}
if we choose a boundary condition approaching to 
a SUSY vacuum labeled by 
$\langle A_1A_2\cdots A_{N_{\rm C}}\rangle $ 
at $y=+\infty$, and to 
a vacuum 
$\langle B_1B_2\cdots B_{N_{\rm C}}\rangle $ 
at $y=-\infty$. 
Since the vacuum condition $\Sigma H^i -H^iM =0$ is satisfied 
at $y=\pm \infty$, 
the second term of the last line of Eq.~(\ref{eq:bogomolnyi}) 
does not contribute. 

The BPS Eqs.~(\ref{BPSeq-Sigma}), 
(\ref{BPSeq-H}) are equivalent to the 
preservation condition of the 
half of the supercharges~\cite{INOS1}.

\section{BPS Wall Solutions}
\label{BPSWS}

The vector multiplet scalar $\Sigma$ can be obtained 
as the sixth component of the six-dimensional 
vector multiplet through the dimensional reduction. 
It is thus natural to combine the fifth component gauge 
field $W_y$ with the vector multiplet scalar to form a 
complexified gauge field $\Sigma + iW_y$. 
Since it is merely a function of $y$, we can eliminate 
it through an $N_{\rm C}\times N_{\rm C}$ invertible 
complex matrix function $S(y)$ 
representing a complexified gauge transformation 
defined as 
\begin{eqnarray}
\Sigma + iW_y \equiv S^{-1}\partial_y S.
\label{def-S}
\end{eqnarray}
Note that this 
differential equation 
determines the function $S(y)$ with 
$N_{\rm C}^2$ arbitrary complex integration constants, 
from which the world-volume symmetry emerges as we see 
later. 
Let us change variables from $H^1,\,H^2$ to 
$N_{\rm C}\times N_{\rm F}$ matrix functions 
$f^1,\,f^2$ by using $S$ 
\begin{eqnarray}
H^1\equiv S^{-1}f^1,\quad H^2\equiv S^\dagger f^2. 
\label{def-f}
\end{eqnarray}
Substituting (\ref{def-S}), (\ref{def-f}) 
to the BPS Eq.~(\ref{BPSeq-H}) 
for $H^i$, we obtain 
\begin{eqnarray}
\partial_y f^1 = f^1 M,\quad \partial _yf^2 = -f^2 M
\end{eqnarray}
which can be easily solved as
\begin{eqnarray}
f^1=H_0^1 \, e^{My},\quad f^2=H^2_0 e^{-My} 
\end{eqnarray}
with the $N_{\rm C}\times N_{\rm F}$ constant 
complex matrices $H_0^1,\,H^2_0$ as 
integration constants, which we call moduli matrices. 
Therefore $H^i$ can be solved completely in terms of $S$ as
\begin{eqnarray}
H^1=S^{-1}H_0^1 e^{My},\quad H^2
=S^\dagger H_0^2 e^{-My}.
\label{sol-H}
\end{eqnarray}
The definitions (\ref{def-S}),\,(\ref{def-f}) show 
that a set 
$(S, H_0^1, H_0^2)$ and 
another set $(S', H_0^1{}', H_0^2{}')$ give 
the same original fields $\Sigma ,\,W_y, H^i$, 
provided they are related by 
\begin{eqnarray}
S' = VS,\quad 
H_0^1{}'=VH_0^1,\quad 
H_0^2{}'
=(V^\dagger )^{-1}H_0^2,
\label{art-sym}
\end{eqnarray} 
where  
$V\in GL (N_{\rm C},{\bf C})$. 
This transformation $V$ defines an equivalence class 
among sets of the matrix function and moduli matrices 
$(S, H_0^1, H_0^2)$ which represents physically 
equivalent results. 
This symmetry  comes from 
the $N_{\rm C}^2$ integration constants in solving 
(\ref{def-S}), and represents 
the redundancy of describing the wall solution 
in terms of $(S, H_0^1, H_0^2)$. 
We call this `world-volume symmetry', 
since this symmetry will be promoted to a 
local gauge symmetry in the world volume of walls 
when we consider the effective action on the walls. 

Since 
$H^2$ vanishes in any SUSY vacuum as given in 
Eq.~(\ref{eq:hyper-vacuum}) as a result of 
non-degenerate masses, 
which we consider here. 
Thus the moduli matrix for domain walls 
$H^2_0$ corresponding to the field $H^2$ also 
vanishes as shown in Appendix C of Ref.~\cite{INOS3}. 
Consequently the field $H^2$ for domain wall solutions 
vanishes identically in the extra dimension.  
Therefore we take
\begin{eqnarray}
 H^2_0=0,\qquad (H^2=0), \qquad H_0 \equiv H^1_0 . 
\end{eqnarray} 

The remaining BPS equations for the vector multiplets 
can be written in terms of the matrix $S$ and the moduli 
matrix $H_0$.   
Since the matrix function $S$ originates from the 
vector multiplet scalars $\Sigma$ and the fifth 
component of the gauge fields $W_y$ as in 
Eq.~(\ref{def-S}), the gauge transformations $U(y)$ 
on the original 
fields $\Sigma ,\,W_y,\,H^1,\,H^2$
\begin{eqnarray}
H^1  &\rightarrow & H^1{}' = U H^1,
\qquad 
H^2  \rightarrow  H^2{}' = U H^2,
\nonumber \\
\Sigma +iW_y &\rightarrow &\Sigma' +iW_y' 
= U\left(\Sigma +iW_y\right)U^\dagger  + U\partial_y U^\dagger ,
\qquad 
\end{eqnarray}
can be obtained by multiplying a unitary matrix 
$U^\dagger(y)$ 
from the right of $S$:  
\begin{eqnarray}
 S\,\rightarrow \,S'=SU^\dagger ,\quad U^\dagger U=1, 
\label{eq:gauge-tr-S}
\end{eqnarray}
without causing any transformations on the moduli 
matrices $H_0$. 
Thus we define 
\begin{eqnarray}
 \Omega  \equiv SS^\dagger , 
\label{def-Omega}
\end{eqnarray}
which is invariant under the gauge transformations 
(\ref{eq:gauge-tr-S}) of the fundamental theory. 
Note that this $\Omega$ is not invariant under the 
world-volume symmetry transformations 
(\ref{art-sym}):
\begin{eqnarray}
 \Omega \ \rightarrow \ \Omega' = V \Omega V^\dagger .
\label{WV-transf-Omega}
\end{eqnarray}
Together with 
the gauge invariant moduli matrix $H_0$, 
the BPS equations (\ref{BPSeq-Sigma}) for vector 
multiplets can be rewritten in the following 
gauge invariant form 
\begin{eqnarray}
\partial_y^2 \Omega -\partial_y \Omega 
\Omega^{-1} \partial_y \Omega = g^2 
\left(c\, \Omega - H_0 \,e^{2My} H_0{}^\dagger 
\right).\label{diff-eq-S}
\end{eqnarray}
Once a solution of $\Omega $ for 
Eq.~(\ref{diff-eq-S}) with a given moduli matrix $H_0$ is 
obtained, 
the matrix $S$ in Eq.~(\ref{def-Omega}) can be determined with a 
suitable gauge choice, and then 
all the quantities, $\Sigma ,\,W_y,\,H^1$ and $H^2$ 
are obtained by Eqs.~(\ref{def-S}) and (\ref{sol-H}). 
Since two boundary conditions are imposed 
at $y=\infty$ and at $y=-\infty$ to the second order 
differential equation (\ref{diff-eq-S}), the number of 
necessary boundary conditions precisely matches to 
obtain the unique solution. 
Therefore there should be no more moduli parameters 
in addition to the moduli 
matrix $H_0$. 
In the limit of infinite gauge coupling, we 
find explicitly that there is no additional moduli. 
We have also analyzed in detail 
the almost analogous nonlinear 
differential equation in the case of the Abelian gauge 
theory at finite gauge coupling and find no additional 
moduli~\cite{IOS1}. 
For the case of non-Abelian gauge theories at 
finite gauge coupling, there is no reason to believe 
a behavior different from the Abelian counterpart.

\section{Moduli Space for Non-Abelian Walls} \label{MSFNAW}

All possible solutions of parallel domain walls 
in the $U(N_{\rm C})$ SUSY gauge theory with 
$N_{\rm F}$ hypermultiplets 
can be constructed once 
the moduli matrix $H_0$ is given. 
The moduli matrix $H_0$ has a redundancy expressed as 
the world-volume symmetry (\ref{art-sym}) : 
$H_0 \sim V H_0$ with 
$V \in GL(N_{\rm C},{\bf C})$. 
We thus find that the moduli space 
denoted by ${\cal M}_{N_{\rm F},N_{\rm C}}$
is homeomorphic to 
the complex Grassmann manifold 
($\tilde N_{\rm C}\equiv N_{\rm F}-N_{\rm C}$) :
\begin{eqnarray}
 {\cal M}_{N_{\rm F},N_{\rm C}}
 \simeq \{H_0 | H_0 \sim V H_0, V \in GL(N_{\rm C},{\bf C})\} 
 \simeq G_{N_{\rm F},N_{\rm C}}
 \simeq {SU(N_{\rm F}) \over 
 SU(N_{\rm C}) \times SU(\tilde N_{\rm C}) \times U(1)}\,.
  \label{Gr}
\end{eqnarray}
This is a {\it compact} (closed) set. 
On the other hand, for instance,  
scattering of two Abelian walls is described by 
a nonlinear sigma model 
on a {\it non-compact} moduli space~\cite{To,IOS1}. 
This fact of the compact 
moduli space consisting of non-compact moduli parameters 
can be consistently understood, if 
we note that the moduli space 
${\cal M}_{N_{\rm F},N_{\rm C}}$ 
includes {\it all BPS topological sectors} 
determined by the different boundary conditions. 
It is decomposed into 
\begin{eqnarray}
 {\cal M}_{N_{\rm F},N_{\rm C}} 
 = \sum_{\rm BPS} 
 {\cal M}^{\langle B_1B_2\cdots B_{N_{\rm C}} \rangle 
\leftarrow 
 \langle A_1A_2\cdots A_{N_{\rm C}} 
\rangle }_{N_{\rm F},N_{\rm C}}, \; 
\end{eqnarray}
where the sum is taken over the BPS sectors. 
Note that it also includes the vacuum states with no walls 
$\langle A_1A_2\cdots A_{N_{\rm C}}\rangle \leftarrow 
\langle A_1A_2 \cdots A_{N_{\rm C}} \rangle$ 
which correspond to 
$N_{\rm F}!/N_{\rm C}! \tilde N_{\rm C}!$ points 
on the moduli space.
Although each sector (except for vacuum states) 
is in general not a closed set,
the total space is compact. 
We call ${\cal M}_{N_{\rm F},N_{\rm C}}$ as 
the ``total moduli space". 
It is useful to rewrite this as a sum over 
the number of walls:
\begin{eqnarray}
 {\cal M}_{N_{\rm F},N_{\rm C}} 
 = \sum_{k=0}^{N_{\rm C}\tilde N_{\rm C}} 
{\cal M}_{N_{\rm F},N_{\rm C}}^k 
 = {\cal M}_{N_{\rm F},N_{\rm C}}^0 
  \oplus {\cal M}_{N_{\rm F},N_{\rm C}}^1 \oplus \cdots 
  \oplus {\cal M}_{N_{\rm F},
N_{\rm C}}^{N_{\rm C}\tilde N_{\rm C}} \,,
 \label{decom}
\end{eqnarray}
with ${\cal M}_{N_{\rm F},N_{\rm C}}^k$ 
the sum of the topological sectors with $k$-walls. 
Since the maximal number of walls is 
$N_{\rm C}\tilde N_{\rm C}$,
${\cal M}_{N_{\rm F},N_{\rm C}}^{N_{\rm C}\tilde N_{\rm C}}$ 
is identical to the maximal topological sector.

This decomposition can be understood as follows. 
Consider a $k$-wall solution  
and imagine a situation such 
that one of the outer-most walls 
goes to spatial infinity. 
We will obtain a ($k-1$)-wall 
configuration in this limit. 
This implies that the $k$-wall sector 
in the moduli space is an open set compactified 
by the moduli space of ($k-1$)-wall sectors on its boundary. 
Continuing this procedure 
we will obtain a single wall configuration. 
Pulling it out to infinity we obtain a vacuum state in the end. 
A vacuum corresponds to a point as 
a boundary of a single wall sector 
in the moduli space.
The $k=0$ sector comprises  
a set of $N_{\rm F}!/N_{\rm C}! \tilde N_{\rm C}!$ points
and does not have any boundary. 
Summing up all sectors,
we thus obtain the total moduli space 
${\cal M}_{N_{\rm F},N_{\rm C}}$ 
as a compact manifold.  
Note again that we include zero-wall sector, 
vacua without any walls. 

The authors in Ref.~\cite{GPTT} constructed a solution of 
the 1/4 BPS equation in a $D=4$, ${\cal N}=2$ SUSY theory 
with $N_{\rm C}=1, N_{\rm F}=2$.
It is a composite soliton made of a wall and vortices 
(strings) ending on it. 
Topological stability of this composite soliton can be understood 
by interpreting vortices as sigma model lumps 
in the effective theory on the wall  
if and only if we consider the total moduli space (\ref{Gr}) 
as the target space instead of the one-wall 
topological sector ${\cal M}_{2,1}^{k=1} \simeq {\bf R} \times S^1$.
This is because the second homotopy group $\pi_2$ 
ensuring the topological stability of lumps
is ${\bf Z}$ for ${\bf C}P^1$
but is trivial for ${\bf R} \times S^1$.
The total moduli space (\ref{Gr}) 
for general $N_{\rm F}$ and/or $N_{\rm C}$ 
provides the topological stability of 
more complicated composite solitons 
which will be discussed in 
Sec.8~\cite{INOS2}.

\section{Infinite Gauge Coupling and Nonlinear Sigma Models}
\label{BPSWSIGC}

The BPS equation (\ref{diff-eq-S}) for the gauge invariant $\Omega$ 
reduces to an algebraic equation 
in the strong gauge coupling limit, given by 
\begin{eqnarray}
 \Omega_{g \to \infty} 
 = (SS^\dagger)_{g \to \infty}  
 = c^{-1}H_0 e^{2My}H_0^{\dagger}. 
  \label{SS-H0}
\end{eqnarray}
Therefore in the infinite gauge coupling 
we do not have to solve the second order differential 
equation for $\Omega$ 
and can explicitly construct wall solutions 
once the the moduli matrix $H_0$ is given. 
Qualitative behavior of walls 
for finite gauge couplings 
is not so different from that
in  infinite gauge couplings. 
This is because the right hand side of 
Eq.~(\ref{diff-eq-S}) tend to zero at 
both spatial infinities even for finite $g$. 
Hence wall solutions for finite $g$ asymptotically 
coincides with those for infinite $g$, 
and they differ only at finite region.  
In fact in \cite{IOS1} 
we have constructed exact wall solutions 
for finite gauge couplings 
and found that their qualitative behavior 
is the same as the infinite gauge coupling cases 
found in the literature~\cite{AT,To
}. 

SUSY gauge theories reduce to 
nonlinear sigma models in general 
in the strong gauge coupling limit 
$g_0 ,\ g \to \infty$.
Since the BPS equations are drastically simplified  
to become solvable in some cases, we often consider this limit. 
In fact the BPS domain walls in theories with 
eight supercharges were first obtained in 
hyper-K\"ahler (HK) nonlinear sigma models~\cite{AT}. 
They have been the only known examples for models with eight 
supercharges~\cite{AT,To
} 
until exact wall solutions 
at finite gauge coupling were found 
recently~\cite{KS,IOS1,IOS2}. 
If hypermultiplets have masses, 
the corresponding nonlinear sigma models 
have potentials, which can be written 
as the square of the tri-holomorphic 
Killing vector on the target manifold~\cite{AF2}. 
These models are called massive HK nonlinear sigma models. 
By this potential most vacua are lifted 
leaving some discrete degenerate points as vacua,
which are characterized by fixed points of the Killing vector. 
In these models interesting composite $1/4$ BPS solitons like 
intersecting walls~\cite{GTT1}, intersecting lumps~\cite{NNS1}  
and composite of wall-lumps~\cite{GPTT,INOS2} 
were constructed.

In our case of $U(N_{\rm C})$ gauge theory with $N_{\rm F})$ 
flavors of fundamental representation, the target space of 
the HK nonlinear sigma model is the cotangent bundle over 
the complex Grassmann manifold~\cite{LR,ANS} 
($\tilde N_{\rm C}\equiv N_{\rm F}-N_{\rm C}$) 
\begin{eqnarray}
\label{T*Gr}
 {\cal M}^{M=0}_{\rm vac} \simeq
 T^* G_{N_{\rm F},N_{\rm C}} \simeq 
 T^* \left[SU(N_{\rm F}) \over 
 SU(N_{\rm C}) \times SU(\tilde N_{\rm C}) \times U(1) 
 \right] \, . 
\end{eqnarray}
From the target manifold (\ref{T*Gr}) one can easily 
see that there exists a duality 
between theories with the same flavor and 
two different gauge groups 
in the case of the infinite gauge coupling~\cite{AP,ANS}: 
$ U(N_{\rm C}) \leftrightarrow U(\tilde N_{\rm C})$. 
This duality holds for the Lagrangian of the 
nonlinear sigma models, and 
leads to the duality between the BPS equations 
of these two theories. 
This duality holds also for the moduli space 
of domain wall configurations. 

We can obtain the effective action on the world volume 
of walls, by promoting the moduli parameters to fields 
on the world volume~\cite{Ma}. 
In the case of infinite gauge coupling, the K\"ahler potential 
of the effective Lagrangian is given by~\cite{INOS3} 
\begin{eqnarray}
 {\cal L}_{\rm walls}^{g \to \infty} 
  = c \int d^4 \theta \int dy \ \log \det \Omega_{g \to \infty} 
  = c \int d^4 \theta \int dy \ \log \det (H_0 e^{2 M y} H_0{}^\dagger) \;. 
 \label{moduli_metric}
\end{eqnarray}

\section{Penetrable and Impenetrable Walls
}
\label{sc:properties}

Let us first take the $U(1)$ gauge theory to 
clarify the physical meaning of the moduli matrix $H_0$, 
which can be parametrized generically by complex moduli 
parameters  $e^{r_A}$ as 
\begin{equation}
H_0=\sqrt{c}(e^{r_1}, e^{r_2}, \cdots, e^{r_{N_F}}). 
\label{eq:moduli-abelian}
\end{equation}
Because of the world-volume symmetry 
(\ref{art-sym}), only the relative magnitude has physical 
meaning: we can fix $e^{r_1}=1$, for instance. 
Then we obtain the hypermultiplet scalars as 
\begin{equation}
H=S^{-1}H_0 e^{My}=S^{-1} 
\sqrt{c}(e^{r_1+m_1y}, \cdots, e^{r_{N_F}+m_{N_F}y}). 
\end{equation}
\begin{figure}[htb] 
\begin{center}
\includegraphics
[width=6cm,height=2.5cm ]
{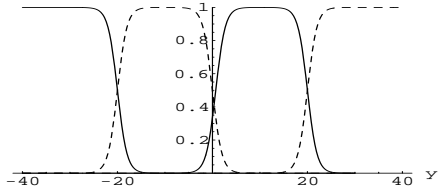}
\caption{Region of rapid change of hypermultiplets indicates the 
positions of walls. }
\label{fig:wall-hyper}
\end{center}
\end{figure}
The relative magnitude 
of different flavors varies as $y$ varies because of the 
nondegenerate masses of the hypermultiplets, 
indicating that the solution approaches 
various vacua successively at various $y$. 
When $A$-th and $A+1$-th flavors 
becomes comparable each other, 
the transition between two adjacent vacua 
occurs. 
This is the location Re$Y_A$ of 
the wall separating the $A$- and $A+1$-th 
vacua, which is determined in terms of  
the relative magnitude of different flavors of 
the moduli matrix elements 
as 
$ Y_A\equiv -{r_A-r_{A+1}\over m_A-m_{A+1}}$. 
The imaginary part Im$Y_A$ of the moduli 
gives the relative phase 
between adjacent vacua. 
The overall normalization is taken care of by the function 
$S^{-1}(y)$. 
Consequently the hypermultiplet scalars exhibit the 
multi-wall behavior as illustrated in 
Fig.~\ref{fig:wall-hyper} 
for the case of $N_{\rm F}=4$ ($N_{\rm C}=1$). 

For multi-walls, we can define the relative distance of 
the walls 
as a difference of two wall positions $R_A=Y_A-Y_{A+1}$. 
If we let $R_A\rightarrow \infty$, we obtain two 
well-separated walls. 
For negative values of the relative distance, 
two walls merge together into a sharp single peak. 
Even in the limit of $R_A\rightarrow -\infty$, we still 
find a single peak with somewhat smaller width, namely 
 a compressed wall. 
We call such a pair of walls as impenetrable. 
In the case of Abelian gauge theories, only the impenetrable 
walls can occur~\cite{To,IOS1}.

The $N_{\rm C}=2, \ N_{\rm F}=4$ case 
is the simplest example 
exhibiting a new feature of 
non-Abelian walls: 
there exists 
a pair of walls whose positions can commute with each other. 
Let us consider the 
topological sector labeled by 
$\langle 13\rangle \leftarrow \langle 24\rangle $. 
A moduli matrix for this sector 
$H_{0\langle 13\leftarrow 24\rangle}$ 
is given by two complex moduli 
parameters $r_2,r_3$ after fixing the world-volume 
symmetry 
\begin{eqnarray}
H_{0\langle 13\leftarrow 24\rangle}=
\sqrt{c}\left( 
\begin{array}{cccc}
1&e^{r_3} &0 &0\\
0&0 & 1& e^{r_2} 
\end{array}
\right).
\end{eqnarray} 
We notice that this moduli matrix factorizes into 
$U(1)$ cases, which is called 
$U(1)$-factorizable, and that 
a solution obtained with this moduli matrix 
is given by 
\begin{eqnarray}
 \Sigma ^0+\Sigma ^3&=&{m_1e^{2m_1y}+m_2e^{2m_2y+2{\rm Re}(r_3)}
\over e^{2m_1y}+e^{2m_2y+2{\rm Re}(r_3)}},\nonumber\\
 \Sigma ^0-\Sigma ^3&=&{m_3e^{2m_3y}+m_4e^{2m_4y+2{\rm Re}(r_2)}
\over e^{2m_3y}+e^{2m_4y+2{\rm Re}(r_2)}}
\end{eqnarray}
and $\Sigma ^1=\Sigma ^2=W_y=0$, and
\unitlength 0.5pt
\begin{figure}[htb]
\begin{center}
\includegraphics[width=4cm,clip]{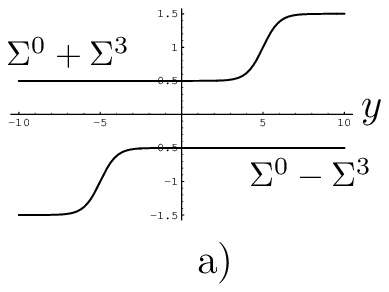}
\hspace{2em}
\includegraphics[width=4cm,clip]{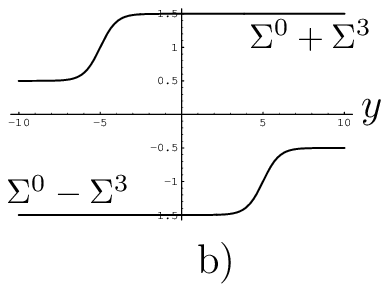}
\vspace{2em}\\
\includegraphics[width=4cm,clip]{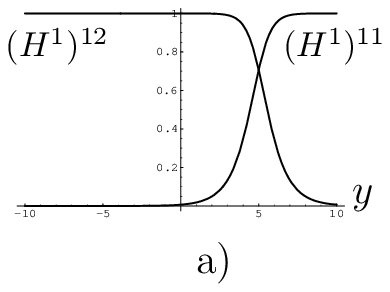}
\hspace{2em}
\includegraphics[width=4cm,clip]{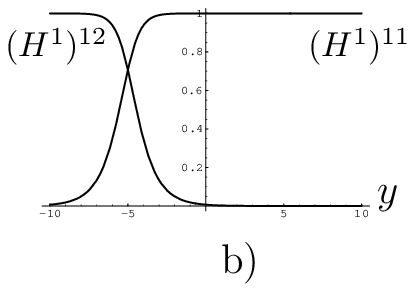}
\vspace{2em}\\
\includegraphics[width=4cm,clip]{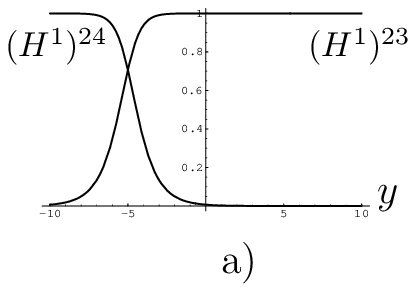}
\hspace{2em}
\includegraphics[width=4cm,clip]{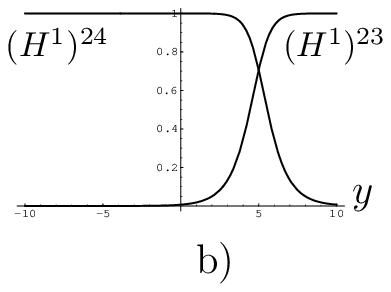}
\end{center}
\caption{
Penetrable walls of $\langle 13 \leftarrow 24 \rangle$ in $N_{\rm C}=2$, 
$N_{\rm F}=4$ case. 
$M={\rm diag.}({3\over 2},{1\over 2},-{1\over 2},-{3\over 2})$. 
a)
 $r_3=5,\,r_2=-5$: 
the intermediate region shows $\langle 23 \rangle$ vacuum. 
b) $r_3=-5,\,r_2=5$, $c=1$: 
the intermediate region shows $\langle 14 \rangle$ vacuum. 
}
\label{commute-walls}
\end{figure}
\unitlength 1pt
 \begin{eqnarray}
 H^1=
 \sqrt{c}\left( 
 \begin{array}{cccc}
 {e^{m_1y}\over \sqrt{e^{2m_1y}+e^{2m_2y+2{\rm Re}(r_3)}}} & 
 {e^{m_2y+r_3}\over \sqrt{e^{2m_1y}+e^{2m_2y+2{\rm Re}(r_3)}}} &0 &0\\
 0&0 & {e^{m_3y}\over \sqrt{e^{2m_3y}+e^{2m_4y+2{\rm Re}(r_2)}}} &{e^{m_4y+r_2}\over \sqrt{e^{2m_3y}+e^{2m_4y+2{\rm Re}(r_2)}}} 
 \end{array}
 \right).\nonumber 
\end{eqnarray}
Profiles of $\Sigma ^0\pm \Sigma ^3$ and $H^1$ 
are illustrated in Fig.~\ref{commute-walls}. 
This solution describes a configuration of double wall. 
In this case, position of the walls are given by 
$y_3= {{\rm Re}(r_3)\over m_1-m_2}$, and 
$y_2={{\rm Re}(r_2)\over m_3-m_4}$. 
In a region $y_2>y_3$, the configuration 
interpolates between vacua $\langle 13\rangle $
and $\langle 24\rangle $ 
through the vicinity of the vacuum $\langle 14\rangle $. 
The wall at $y_2$ ($y_3$) interpolates between 
vacua $\langle 13\rangle $ and $\langle 14\rangle $ 
($\langle 24\rangle $ and $\langle 14\rangle $). 
On the other hand, in a region $y_3>y_2$ 
the wall at $y_3$ ($y_2$) interpolates 
as $\langle 13\rangle \leftarrow \langle 23\rangle $ 
($\langle 23\rangle \leftarrow \langle 24\rangle $). 
Note that  while 
intermediate vacua are different from one region $(y_3 > y_2)$
 to the other $(y_3 < y_2)$, 
the two parameters $y_2,y_3$ retain the physical meaning as 
positions of the walls. 
The wall represented 
by the position $y_2$ changes 
flavors of non-vanishing hypermultiplet scalar 
from 4 to 3 and changes the value of $\Sigma ^0-\Sigma ^3$, 
while the wall at $y_3$ changes flavors from 2 to 1 and 
changes $\Sigma ^0+\Sigma ^3$.  
Thus, it is quite natural to identify the wall 
represented by the same parameter,  
although interpolated vacua are
different. 
With this identification, we find that these 
two walls maintain their identity after 
passing through each other. 
Therefore we call the pair of walls as penetrable 
walls~\cite{INOS3}. 

We also found that these properties of non-Abelian walls 
can be conveniently summarized by means of walls 
operators acting on the moduli space~\cite{INOS3}.

\section{Vortices with 
Walls }
\label{sc:vortex}

If we make the moduli matrix $H_0$ 
to depend on 
the world volume coordinates, the wall location 
should depend on the position in the world volume. 
Then the walls will in general be curved and a magnetic 
field is generated. 
If we have a zero in $H_0$, it will give us a 
spike-like behavior of the wall, which becomes a 
vortex ending at the wall~\cite{GPTT}. 
Similarly, if we allow an exponential dependence on 
the world volume coordinates for the moduli 
matrix elements, the wall can tilt. 
A vortex alone preserves a 
different combination of $1/2$ of supercharges compared to the 
wall perpendicular to the vortex. 
In fact we have found that the 
addition of vortex perpendicular to the walls 
can preserve $1/2$ of 
the surviving SUSY on the world volume 
of the wall. 
Consequently the combined configuration preserves 
$1/4$ of the original SUSY. 

We assume the vortices depend on $x^1, x^2$ and walls 
on $x^3$ 
(the combined configuration depends on 
$x^m\equiv x^1, x^2, x^3$ : 
co-dimension three) and assume Poincar\'e invariance 
in $x^0,x^4$. 
Then we obtain $W_0=W_4=0$. 
Using the arguments for the walls, we take $H^2=0$. 
Requiring the SUSY transformation of fermions to vanish 
along the above $1/4$ SUSY directions, 
a set of 1/4 BPS equations 
is obtained. 
The wall BPS equation has 
an additional contribution of vortex magnetic field 
$F_{12}$ 
\begin{eqnarray}
{\cal D}_3 \Sigma
&=&{g^2\over 2}
\left(c{\bf 1}_{N_{\rm C}}-H^1H^1{}^\dagger\right)
+F_{12} ,
\quad {\cal D}_3 H^1 
=
-\Sigma H^1 +H M,
\label{1/4BPSeq-1}
\end{eqnarray}
which is supplemented by the BPS equations for vortices 
\begin{eqnarray}
0&=&{\cal D}_1 H^1 +i{\cal D}_2 H^1,
\quad 
0
=
F_{23}
-{\cal D}_1 \Sigma,\quad 
0=F_{31}
-{\cal D}_2 
\Sigma. 
\label{1/4BPSeq}
\end{eqnarray}
We have found that this set of $1/4$ BPS equations 
allows the recently found monopole in the Higgs 
phase~\cite{mono-Higgs}. 

We obtain 
the BPS bound of the energy density 
$\cal E$ as 
${\cal E}\geq t_{\rm w}+t_{\rm v}+t_{\rm m}+\partial _mJ_m$ 
with $t_{\rm w},\,t_{\rm v}$ and $t_{\rm m}$ 
the energy densities for walls, vortices and monopoles, 
and the correction term $J_m$, which does not contribute 
for individual walls and vortices 
\begin{eqnarray}
 t_{\rm w}&=&c\partial _3 {\rm Tr}\Sigma,\quad 
 t_{\rm v}=-c{\rm Tr}\!  F_{12},\quad 
 t_{\rm m}
=
\frac{2}{g^2}\partial_m {\rm Tr}
( \frac12 \epsilon^{mnl} \! F_{nl}\Sigma ), 
 \label{energy-den}
\end{eqnarray}
\begin{eqnarray}
 J_1&=&{\rm Re}\left(-i{\rm Tr}(H_i^\dagger {\cal D}_2H^i)\right),\, 
J_2={\rm Re}\left(i{\rm Tr}(H_i^\dagger {\cal D}_1H^i)\right),\nonumber\\
J_3&=&-{\rm Tr}(H^\dagger _i(\Sigma -M)H^i).
\end{eqnarray}
Let us note that the magnetic flux from our monopole is 
measured in terms of the dual field strength multiplied 
(projected) by the Higgs field 
$\frac12 \epsilon^{mnl} \! F_{nl}\Sigma$, 
as is usual to obtain the $U(1)$ field strength for 
the monopole in the Higgs phase~\cite{mono-Higgs}. 

It is crucial to observe that Eq.~(\ref{1/4BPSeq}) 
guarantees the integrability condition 
$[{\cal D}_1+i{\cal D}_2,\,{\cal D}_3+\Sigma]
=[\partial_1+i \partial_2,\,\partial_3]=0$. 
Therefore we can introduce an $N_{\rm C}\times N_{\rm C}$ 
invertible complex matrix function 
$S(x^m)\in GL(N_{\rm C},\mathbf{C})$ 
defined by 
\begin{eqnarray}
\Sigma + iW_3 &\equiv& S^{-1}\partial_3 S,\quad 
\left(({\cal D}_3+\Sigma)S^{-1}=0\right),
\label{def-S-3} \\
W_1+iW_2&\equiv &-2iS^{-1}\bar \partial S,\, 
\left(({\cal D}_1+i{\cal D}_2)S^{-1}=0\right) , 
\quad \label{def-S-bar}
\end{eqnarray}
where $z \equiv x^1 + i x^2$, and 
$\bar \partial\equiv \partial/\partial z^*$. 
With (\ref{def-S-3}) and (\ref{def-S-bar}), 
other Eqs.~(\ref{1/4BPSeq}) are automatically satisfied, 
or are easily solved without any assumptions by 
\begin{eqnarray}
 H^1 = S^{-1}(z,z^*, x^3)H_0(z) e^{M x^3}.
\label{sol-H-vortex}
\end{eqnarray}
Here $H_0(z)$ is an arbitrary $N_{\rm C} \times N_{\rm F}$ 
matrix whose elements are arbitrary holomorphic functions 
of $z$. 
We call it  ``moduli matrix''.

Defining an $N_{\rm C}\times N_{\rm C}$ 
Hermitian matrix $\Omega \equiv SS^\dagger $, 
invariant under the $U(N_{\rm C})$ gauge transformations, 
we obtain the remaining BPS equation in Eq.~(\ref{1/4BPSeq}) 
\begin{eqnarray}
4\partial \bar \partial \Omega 
-4(\partial \Omega )\Omega ^{-1}
(\bar \partial \Omega )
+\partial _3^2\Omega -(\partial _3\Omega )
\Omega ^{-1}(\partial _3\Omega )
= g^2 
\left(c\, \Omega - H_0 \,e^{2My} H_0{}^\dagger 
\right). \label{diff-eq-S-vortex}
\end{eqnarray}

\begin{figure}[htb]
\includegraphics[width=4.0cm]{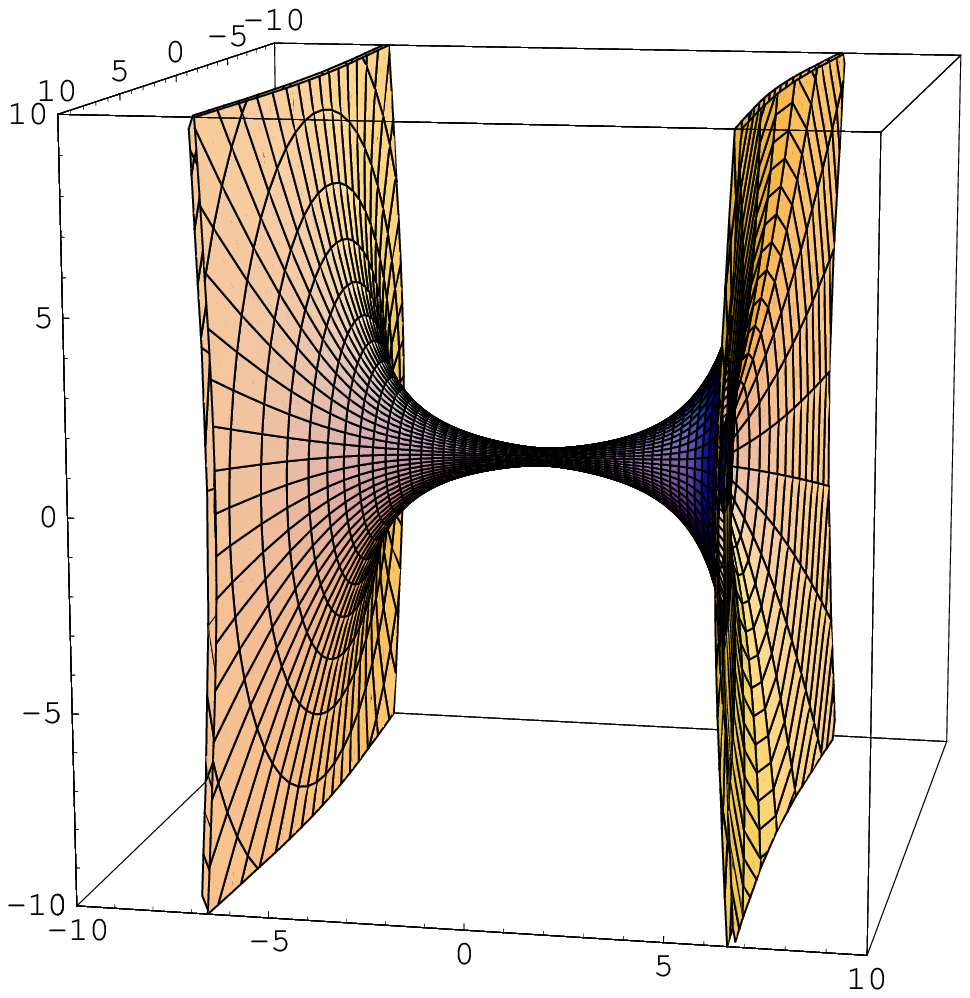}
\put(-75,-2){$x^3$}
\put(-122,60){$x^1$}
\put(-105,110){$x^2$}
\put(-60,-7){a)}
\includegraphics[width=7.5cm]{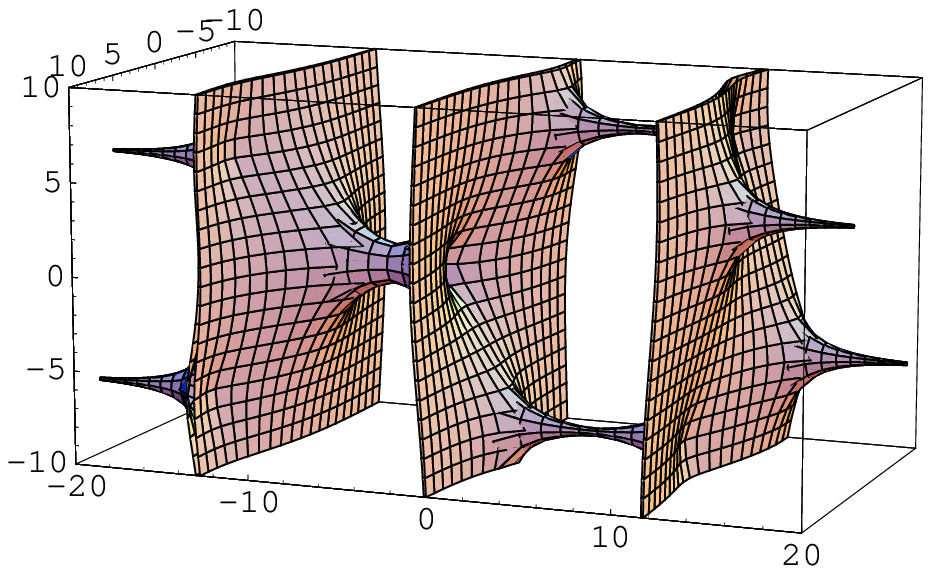}
\put(-140,0){$x^3$}
\put(-205,65){$x^1$}
\put(-185,118){$x^2$}
\put(-100,-7){b)}
\vspace{1cm}
\includegraphics[width=4.0cm]{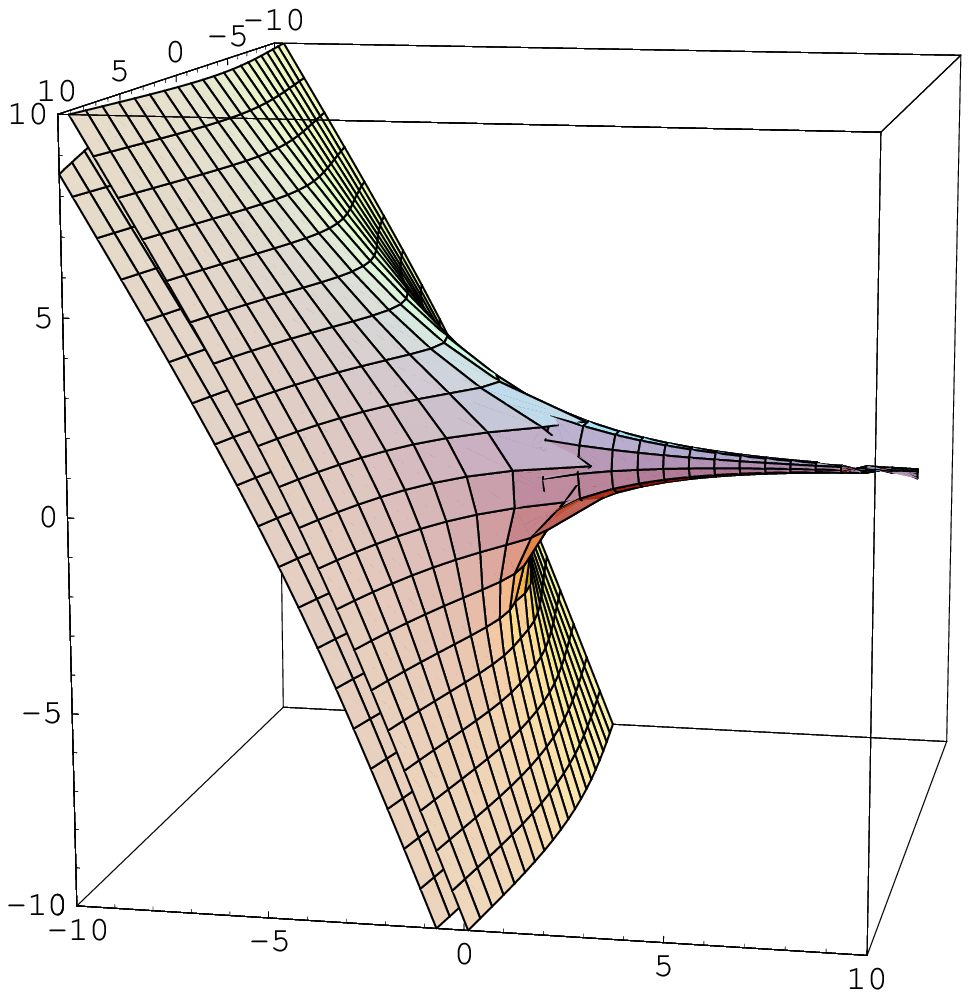}
\put(-75,-2){$x^3$}
\put(-118,60){$x^1$}
\put(-105,110){$x^2$}
\put(-60,-7){c)}
\caption{\label{fig:wormhole} Surfaces defined by the 
same energy density. 
Note that there are two surfaces with the same energy 
for each wall.
a) A vortex stretched between walls with 
$H_0(z)e^{Mx^3}=\sqrt{c}(e^{x^3},ze^{4},e^{-x^3})$. 
b)  Multi-vortices between multi-walls 
with 
$H_0(z)e^{Mx^3}=
\sqrt{c}((z-4-2i)(z+5+8i)e^{3/2x^3},
(z+8-i)(z-7+6i)e^{1/2x^3+12},
z^2e^{-1/2x^3+12},(z-6-5i)(z+6-7i)e^{-3/2x^3})$. 
c) A vortex 
attached to a tilted wall with 
$H_0(z)e^{Mx^3}=\sqrt{c}(z^2e^{x^3},e^{-1/2z})$.
}
\end{figure}

Eqs.~(\ref{def-S-3})--(\ref{diff-eq-S-vortex}) 
determine a map 
from our moduli matrix $H_0(z)$ 
to all possible $1/4$ BPS solutions in three-dimensional 
configuration space. 
Therefore we find that the moduli space of the $1/4$ BPS 
solutions is the space of all the holomorphic maps 
from a complex plane to the wall moduli space, 
the deformed complex Grassmann manifold 
$ G_{N_{\rm F},N_{\rm C}}=
{SU(N_{\rm F}) \over 
 SU(N_{\rm C}) \times SU(\tilde N_{\rm C}) \times U(1)}$ 
in Eq.(\ref{Gr}). 

We can obtain exact solutions in the limit of 
infinite gauge coupling, since the BPS 
equation (\ref{diff-eq-S-vortex}) 
can be solved algebraically: 
$c\, \Omega = H_0 \,e^{2My} H_0{}^\dagger$. 
To illustrate the power of our method, let us take 
the simplest case of $U(1)$ gauge theory with $N_{\rm F}$ 
flavors. 
The moduli vector 
$H_0(z)=\sqrt{c}\left(f^1(z),\,\dots,
f^{N_{\rm F}}(z)\right)$ gives 
$
 \Omega =\sum_{A=1}^{N_{\rm F}}|f^A(z)|^2e^{2m_Ax^3}$. 
Since the position of the $A$-th wall separating $A$-th and 
$(A+1)$-th vacua is given by 
$
x_A^3(z)=\left(\log|f_{A+1}(z)|-\log|f_A(z)|\right)/(m_A-m_{A+1})
$, $z$ dependence indicates a curved wall. 
Holomorphy of the moduli matrix allows only zeros or exponential 
dependence (entire function). 
If we choose zeros in 
the moduli matrix 
\begin{eqnarray}
 f^A(z)=f^A_0\prod_{\alpha }(z-z^A_{\alpha })^{k^A_{\alpha }}, 
  \quad k^A_\alpha \in {\bf Z}_+, 
\end{eqnarray}
we obtain vortices of vorticity $k_{\alpha}^A$ 
at $z=z^A_{\alpha}$ in the $A$-th vacuum. 
A single vortex stretching between two walls is shown 
in Fig.~\ref{fig:wormhole}a), where logarithmic bending 
of the wall is visible towards $|z|\rightarrow \infty $. 
We can avoid the logarithmic bending by 
requiring 
$ k=\sum_{\alpha }k^A_\alpha$ to be common to walls, 
as illustrated in Fig.~\ref{fig:wormhole}b). 
A monopole in the Higgs phase 
was found recently as a kink on the world volume 
of vortex in non-Abelian gauge theories~\cite{mono-Higgs}. 
Even in the $U(1)$ gauge theory, a similar kink 
occurs on the world volume of vortex. 
The right (left) wall separates the first (third) 
and the second vacua. 
The vortex can be regarded as a spike-like bending 
of the wall. 
Therefore there must be a transition from the first 
vacuum to the third vacuum when we go from right to 
left of the vortex. 
In the middle of the vortex in Fig~\ref{fig:wormhole}a), 
there must be a kink, 
analogous to the monopole in the Higgs phase. 
Because of $1/g^2$ factor, 
the energy density of monopoles $t_{\rm m}$ vanish 
in the limit of infinite gauge coupling. 
The monopole charge $g^2t_{\rm m}$ is, however, 
finite as a kink on the vortex, precisely analogous 
to the non-Abelian case~\cite{mono-Higgs}. 
In a simple example of a vortex with winding number $k$ 
stretched between two walls, we obtain 
the monopole charge 
\begin{eqnarray}
\int _Vd^3x g^2 t_{\rm m}=-\pi |m_A-m_B|k.
\end{eqnarray}
Let us stress that our monopole in the Higgs phase should 
give non-vanishing contribution to the energy density once 
gauge coupling becomes finite.

If we allow an exponential function of $z$, 
such as $e^{m z}$, 
somewhere in the moduli matrix $H_0(z)$, 
the corresponding wall is no longer perpendicular to 
the $x^3$-axis. 
If we choose $H_0(z)$, for instance, 
\begin{eqnarray}
H_0e^{Mx^3}\!=\! 
\sqrt{c}(e^{m_1x^3+\tilde m_1 z},\,e^{m_2x^3+\tilde m_2 z}),
\,\,\tilde m_{1, 2} \in {\mathbf C}, 
\end{eqnarray}
we find that the wall position is expressed as
$m_1x^3+{\rm Re}(\tilde m_1 z)=$ 
$m_2x^3+{\rm Re}(\tilde m_2 z)$. 
Moreover we find 
the magnetic field strength 
flows down along the tilted wall 
to negative infinity of $x^3$. 
If vortices are present, they are no longer perpendicular 
to such tilted walls as illustrated 
in Fig.~\ref{fig:wormhole}c).
This configuration offers a field theoretical model 
of the string ending on the D-brane with a magnetic 
flux~\cite{HH}.

\section{Acknowledgements}

The authors thank Koji Hashimoto and 
David Tong for a useful discussion. 
This work is supported in part by Grant-in-Aid for Scientific 
Research from the Ministry of Education, Culture, Sports, 
Science and Technology No.13640269 
and 16028203 for the priority area ``origin of mass''. 
K.O. and M.N. are 
supported by Japan Society for the Promotion 
of Science under the Post-doctoral Research Program. 
Y.I. gratefully acknowledges 
support from a 21st Century COE Program at 
Tokyo Tech ``Nanometer-Scale Quantum Physics'' by the 
Ministry of Education, Culture, Sports, Science 
and Technology.

\bibliographystyle{plain}

\end{document}